\documentclass[12pt,a4paper]{article}
\usepackage{amsmath}
\usepackage{xspace}
\usepackage{epsfig}
\usepackage{rotating}
\usepackage{dcolumn}
\usepackage{url}
\usepackage{hyperref}
\hypersetup{ pdfborder=0 0 0, urlbordercolor=1 0 0 }
\usepackage{cite}

\newcommand{\etal}{{\it et al.}}

\newcommand{\Bs}{{B}_s^0}

\usepackage{graphicx}
\usepackage{color}
\usepackage{colortbl}

\begin{document}


\begin{titlepage}
\belowpdfbookmark{Title page}{title}

\pagenumbering{roman}
\vspace*{-1.5cm}
\centerline{\large EUROPEAN ORGANIZATION FOR NUCLEAR RESEARCH (CERN)}
\vspace*{1.5cm}
\hspace*{-5mm}\begin{tabular*}{16cm}{lc@{\extracolsep{\fill}}r}
\vspace*{-12mm}\mbox{\!\!\!\epsfig{figure=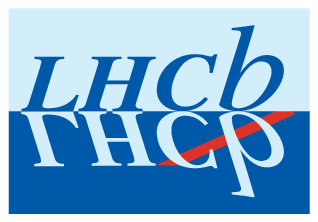,width=.12\textwidth}}& & \\
&& CERN-PH-EP-2011-011\\
&&1 February 2011 \\
\end{tabular*}
\vspace*{4cm}
\begin{center}
{\bf\huge\boldmath {First observation of $\Bs\to J/\psi f_0(980)$ decays}\\
}
\vspace*{2cm}
\normalsize {
The LHCb Collaboration\footnote{Authors are listed on the following pages.}
}
\end{center}
\vspace{\fill}
\centerline{\bf Abstract}
\vspace*{5mm}\noindent
Using data collected with the LHCb detector in proton-proton collisions at a centre-of-mass energy of 7 TeV,  the hadronic decay  $\Bs\to J/\psi f_0(980)$ is observed. This CP eigenstate mode could be used  
to measure mixing-induced CP violation in the $\Bs$ system.
Using a fit to the $\pi^+\pi^-$ mass spectrum with interfering resonances gives $R_{f_0/\phi}\equiv\frac{\Gamma(B_s^0\to J/\psi f_0,~f_0\to\pi^+\pi^-)}{\Gamma(B_s^0\to J/\psi \phi,~\phi\to K^+K^-)}=0.252^{+0.046+0.027}_{-0.032-0.033}$. 
In the interval $\pm$90 MeV around 980 MeV, corresponding to approximately two full $f_0$ widths we also find
$R'\equiv\frac{\Gamma \left(B_s^0\to J/\psi\pi^+\pi^-,~\left| m(\pi^+\pi^-)-980~{\rm MeV}\right|<90~{\rm MeV}\right)}{\Gamma(B_s^0\to J/\psi \phi,~\phi\to K^+K^-)}=0.162\pm 0.022\pm 0.016$, where
in both cases  the uncertainties are statistical and systematic, respectively.

\vspace*{1.cm}

\vspace*{1.cm}
\center{To be published in Physics Letters B}
\vspace{\fill}

\end{titlepage}

\setcounter{page}{2}

\belowpdfbookmark{LHCb author list}{authors}
\centerline{\Large The LHCb Collaboration}
\begin{flushleft}
\small
R.~Aaij$^{23}$, 
B.~Adeva$^{36}$, 
M.~Adinolfi$^{42}$, 
C.~Adrover$^{6}$, 
A.~Affolder$^{48}$, 
Z.~Ajaltouni$^{5}$, 
J.~Albrecht$^{37}$, 
F.~Alessio$^{6,37}$, 
M.~Alexander$^{47}$, 
P.~Alvarez~Cartelle$^{36}$
% ...
\bigskip\newline{\it\footnotesize
% ...
$ ^{5}$Clermont Universit\'{e}, Universit\'{e} Blaise Pascal, CNRS/IN2P3, LPC, Clermont-Ferrand, France\\
$ ^{6}$CPPM, Aix-Marseille Universit\'{e}, CNRS/IN2P3, Marseille, France\\
% ...
$ ^{23}$Nikhef National Institute for Subatomic Physics, Amsterdam, Netherlands\\
% ...
$ ^{36}$Universidad de Santiago de Compostela, Santiago de Compostela, Spain\\
$ ^{37}$European Organization for Nuclear Research (CERN), Geneva, Switzerland\\
% ...
$ ^{42}$H.H. Wills Physics Laboratory, University of Bristol, Bristol, United Kingdom\\
% ...
$ ^{47}$School of Physics and Astronomy, University of Glasgow, Glasgow, United Kingdom\\
$ ^{48}$Oliver Lodge Laboratory, University of Liverpool, Liverpool, United Kingdom\\
% ...
}
\end{flushleft}

\cleardoublepage
\setcounter{page}{1}
\pagenumbering{arabic}

\newpage

\section{Introduction}

In $B_s^0$ decays some final states can be reached either by a direct decay amplitude or via a mixing amplitude. For the case of $B_s^0\to J/\psi\phi$ decays, the interference between these two amplitudes allows observation of a CP violating phase. In the Standard Model (SM) this phase is $-2\beta_s=-0.036^{+0.0020}_{-0.0016}$ radians, where $\beta_s=\arg\left(-V_{ts}V_{tb}^*/V_{cs}V_{cb}^*\right)$, and the $V_{ij}$ are CKM matrix elements \cite{Charles}.   This is about 20 times smaller in magnitude than the measured value of the corresponding phase $2\beta$ in ${B}^0$ mixing. Being small, this phase can be drastically increased by the presence of new particles beyond the SM. Thus, measuring $\beta_s$ is an important probe of new physics.

Attempts to determine $\beta_s$ have been made by the CDF and D0 experiments at the Tevatron using the $\Bs\to J/\psi\phi$ decay mode \cite{Tevatron0}. While initial results hinted at possible large deviations from the SM, recent measurements are more consistent \cite{TevatronCDF,TevatronD0}. However, the Tevatron limits are still not very constraining. Since the final state consists of two spin-1 particles, it is not a CP eigenstate. While it is well known that CP violation can be measured using angular analyses \cite{transversity}, this requires more events to gain similar sensitivities to those obtained if the decay proceeds via only  CP-even or CP-odd channels. In Ref.~\cite{Stone-Zhang} it is argued that in the case of $J/\psi\phi$ the analysis is complicated by the presence of an S-wave $K^+K^-$ system interfering with the $\phi$ that must be taken into account, and that this S-wave would also manifest itself by the appearance of $f_0(980)\to\pi^+\pi^-$ decays.  This decay $\Bs\to J/\psi f_0(980)$ is to a single CP-odd eigenstate and does not require an angular analysis. Its CP violating phase in the Standard Model is  $-2\beta_s$ (up to corrections due to higher order diagrams). In what follows, we use the notation $f_0$ to refer to the $f_0(980)$ state.

By comparing  $D_s^+\to f_0 \pi^+$ decays where the $f_0$ was detected in both $K^+K^-$ and $\pi^+\pi^-$ modes it was predicted that \cite{Stone-Zhang}
\begin{equation}
R_{f_0/\phi}\equiv\frac{\Gamma(B_s^0\to J/\psi f_0,~f_0\to \pi^+\pi^-)}{\Gamma(B_s^0\to J/\psi \phi,~\phi\to K^+K^-)}
\approx 20\%.
\end{equation}
A decay rate at this level would make these events very useful for measuring $\beta_s$ if backgrounds are not too large. 

\begin{figure}[hbt]
\centering
\includegraphics[width=3.in]{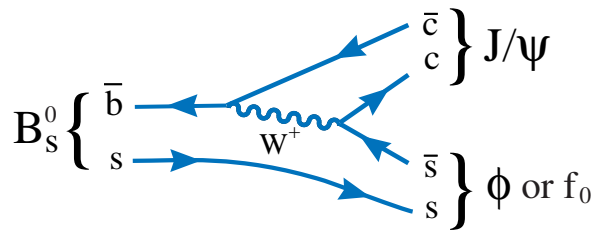}
\caption{Decay diagram for $B_s^0\to J/\psi (f_0{\rm ~or~}\phi)$ decays.
 } \label{psi_f0-phi}
\end{figure}

The dominant decay diagram for these processes is shown in Fig.~\ref{psi_f0-phi}.
It is important to realize that the $s\overline{s}$ system accompanying the $J/\psi$ is an isospin singlet (isoscalar), and thus cannot produce  a single meson that is anything but isospin zero. Thus, for example, in this spectator model production of a $\rho^0$ meson is forbidden. The dominant low mass  isoscalar resonance decaying into $\pi^+\pi^-$ is the $f_0(980)$ but other higher mass objects are possible.

Although the $f_0$ mass is relatively well estimated at 980$\pm$10 MeV (we use units with $c = 1$) by the PDG, the width is poorly known. Its measurement appears to depend on the final state, and is complicated by the opening of the $KK$ channel close to the pole; the PDG estimates 40$-$100 MeV \cite{PDG}. Recently CLEO measured these properties in the semileptonic decay $D_s^+\to f_0 e^+\nu$, where hadronic effects are greatly reduced, determining a width of $(91^{+30}_{-22}\pm 3)$ MeV \cite{CLEO-f0-semi}.

\section{Data sample and analysis requirements}
We use a data sample
of approximately 33 pb$^{-1}$ collected with the LHCb detector in 2010 \cite{LHCb-det}. The detector elements are placed along the beam line of the LHC starting with the Vertex Locator (VELO), a silicon strip device that surrounds the proton-proton interaction region and is positioned 8 mm from the beam during collisions. It provides precise locations for primary $pp$ interaction vertices, the locations of decays of long-lived particles, and contributes to the measurement of track momenta.  Other devices used to measure track momenta comprise a large area silicon strip detector (TT) located in front of a 3.7 Tm dipole magnet, and a combination of
silicon strip detectors (IT) and straw drift chambers (OT) placed behind. Two Ring Imaging Cherenkov (RICH) detectors are used to identify charged hadrons. Further downstream an Electromagnetic Calorimeter (ECAL) is used for photon detection and electron identification, followed by a Hadron Calorimeter (HCAL), and  a system consisting of alternating layers of iron and chambers (MWPC and triple-GEM) that distinguishes muons from hadrons (MUON). The ECAL, MUON, and HCAL provide the capability of first-level hardware triggering.

\begin{figure}[!hbt]
\centering
\includegraphics[width=3.5in]{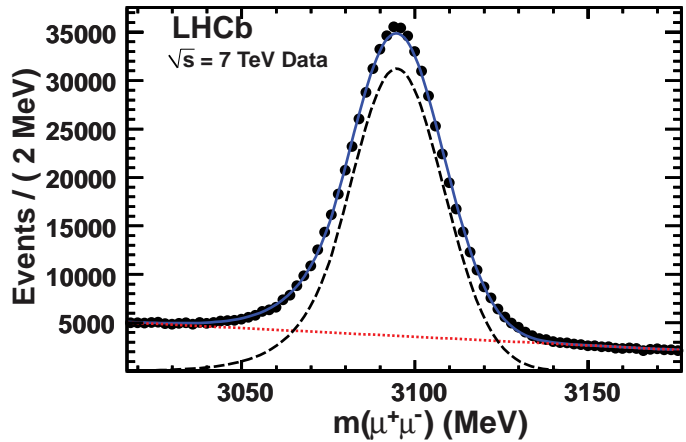}
\caption{The $\mu^+\mu^-$ invariant mass for candidates satisfying the trigger and analysis requirements and having $t_z>0.5$ ps. The data points are shown as circles; the error bars are smaller than the circle radii. The dashed line shows the Crystal Ball signal function \cite{CB}, the dotted line the background and the solid line the sum.
} \label{mjpsi_data}
\end{figure}

This analysis is restricted to events accepted by a $J/\psi\to\mu^+\mu^-$ trigger.
Subsequent analysis selection criteria are applied that serve to reject background, yet preserve high efficiencies on both the $J/\psi \pi^+\pi^-$ and $J/\psi K^+K^-$ final states, as determined by Monte Carlo events generated using PYTHIA \cite{Pythia}, and LHCb detector simulation based on GEANT4 \cite{GEANT4}. Tracks are reconstructed as described in Ref.~\cite{LHCb-det}.
To be considered as a $J/\psi\to\mu^+\mu^-$ candidate opposite sign tracks are required to have  transverse momentum, $p_{\rm T}$, greater than 500 MeV, be identified as muons, and 
form a common vertex with fit  $\chi^2$ per number of degrees of freedom (ndof) less than 11. The $\mu^+\mu^-$ invariant mass distribution is shown in Fig.~\ref{mjpsi_data} with an additional requirement, used only for this plot, that the pseudo proper-time, $t_z$, be greater than 0.5 ps, where $t_z$ is the distance that the $J/\psi$ candidate travels downstream parallel to the beam, along $z$, times the known $J/\psi$ mass divided by the $z$ component of the candidate's momentum. The data are fit with a Crystal Ball signal function \cite{CB}  to account for the radiative tail towards low mass, and a linear background function. There are 549,000$\pm$1100 $J/\psi$ signal events in the entire mass range.
For subsequent use only candidates within $\pm$48 MeV of the known $J/\psi$ mass are selected. 

Pion and kaon candidates are selected if they are inconsistent with having been produced at the closest primary vertex. The impact parameter (IP) is the minimum distance of approach of the track with respect to the primary vertex. We require that the $\chi^2$ formed by using the hypothesis that the IP is equal to zero be $>9$ for each track. For further consideration these tracks must be positively identified in the RICH system. Particles forming opposite-sign di-pion candidates must have their scalar sum $p_{\rm T}> 900$ MeV, while those forming opposite-sign di-kaon candidates must have their vector sum $p_{\rm T}> 1000$ MeV, and have an invariant mass within $\pm$20 MeV of the $\phi$ mass.

To select $B_s^0$ candidates we further require that the two pions or kaons form a vertex with a $\chi^2< 10$, that they form a candidate $B_s^0$ vertex with the $J/\psi$ where the vertex fit $\chi^2$/ndof $<5$, and that this $B_s^0$ candidate points to the primary vertex at an angle not different from its momentum direction by more than 0.68$^{\circ}$.

Simulations are used  to evaluate our detection efficiencies. For the $J/\psi\phi$ final state we use the measured decay parameters from CDF \cite{TevatronCDF}.
The $J/\psi f_0$ final state is simulated using full longitudinal polarization of the $J/\psi$ meson.
The efficiencies of having all four decay tracks in the geometric acceptance and satisfying the trigger, track reconstruction and data selection requirements are (1.471$\pm$0.024)\% for $J/\psi f_0$, requiring the $\pi^+\pi^-$ invariant mass be within $\pm$500 MeV of 980 MeV, and (1.454$\pm$0.021)\% for $J/\psi\phi$, having the $K^+K^-$ invariant mass be within $\pm$20 MeV of the $\phi$ mass. The uncertainties on the efficiency estimates are statistical only.


\section{Results}

\begin{figure}[hbt]
\centering
\includegraphics[width=5.in]{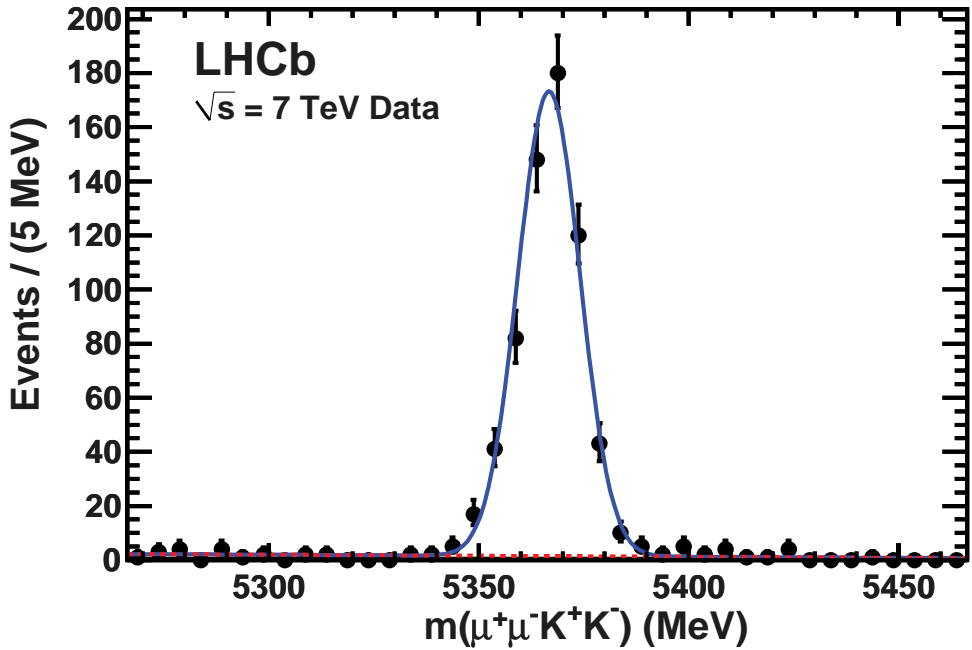}
\caption{The invariant mass of $J/\psi K^+K^-$ combinations when the $K^+K^-$ pair is required to be with $\pm$20 MeV of the $\phi$ mass. The data have been fit with a Gaussian signal function whose mass and width are allowed to float and linear background function shown as a dashed line. The solid curve shows the sum. } 
\label{Bs2Jpsiphi_cb}
\end{figure}

The $J/\psi K^+K^-$ invariant mass distribution is shown in Fig.~\ref{Bs2Jpsiphi_cb}. The di-muon invariant mass has been constrained to have the known value of the $J/\psi$ mass; this is done for all subsequent $\Bs$ invariant mass distributions. The data are fit with a Gaussian signal function and a linear background function. The fit gives a $\Bs$ mass of 5366.7$\pm$0.4 MeV, a width of 7.4 MeV r.m.s., and a yield of 635$\pm$26 events. 

\begin{figure}[hbt]
\centering
\includegraphics[width=4.5in]{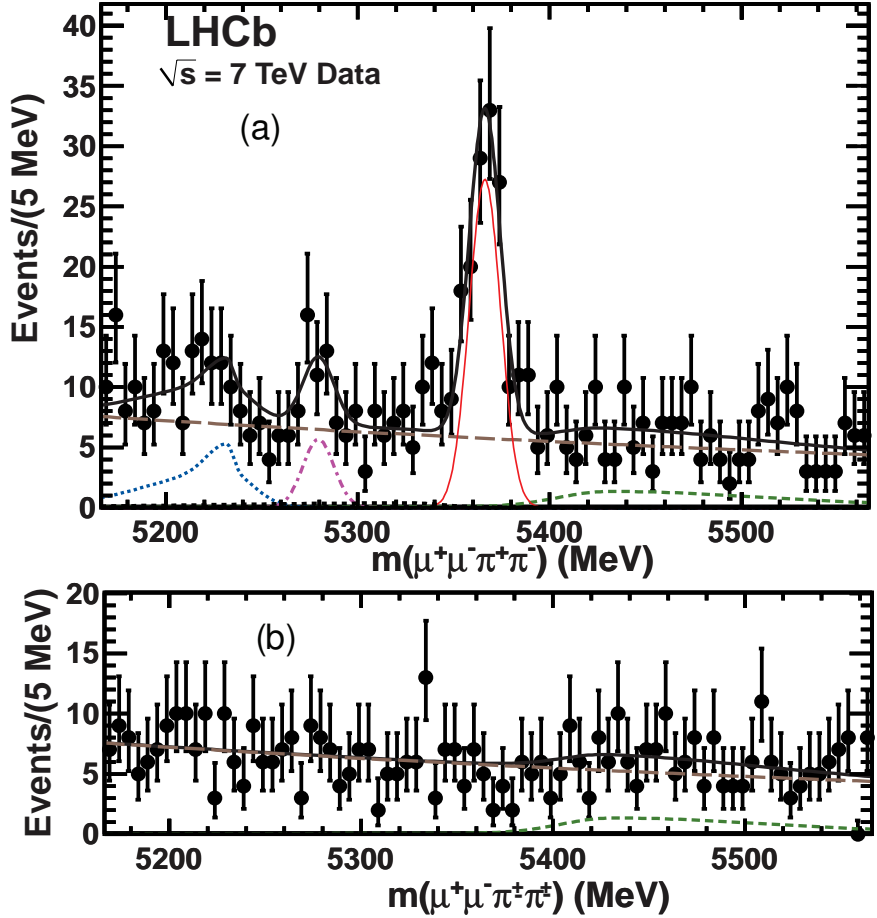}
\caption{(a) The invariant mass of $J/\psi\pi^+\pi^-$ combinations when the $\pi^+\pi^-$ pair is required to be with $\pm$90 MeV of the $f_0(980)$ mass. The data have been fit with a signal Gaussian and several background functions. The thin (red) solid curve shows the signal, the long-dashed (brown) curve  the combinatorial background, the dashed (green) curve  the $B^+\to J/\psi K^+(\pi^+) $ background, the dotted (blue) curve  the ${B}^0\to J/\psi K^{*0}$ background, the dash-dot curve (purple) the ${B}^0\to J/\psi \pi^+\pi^-$ background, the barely visible dotted curve  (black) the sum of $B_s^0\to J/\psi \eta'$ and $J/\psi\phi$ backgrounds, and the thick-solid (black) curve the total. (b) The same as above but for like-sign di-pion combinations.
 } \label{fit_bsmass}

\end{figure}

Initially, to search for a $f_0(980)$ signal we restrict ourselves to an interval of  $\pm$90 MeV around the $f_0$ mass, approximately two full $f_0$ widths \cite{CLEO-f0-semi}. 
The $B_s^0$ candidate invariant mass distribution for selected $J/\psi\pi^+\pi^-$ combinations is shown in Fig.~\ref{fit_bsmass}. The signal is fit with a Gaussian whose mean and width are allowed to float.  
We also include a background component due to ${B}^0\to J/\psi \pi^+\pi^-$ that is taken to be Gaussian, with mass allowed to float in the fit, but whose width is constrained to be the same as the $B_s^0$ signal. 
Other components in the fit are ${B}^0\to J/\psi K^{*0}$,  combinatorial background taken to have an exponential shape,  $B^+\to J/\psi K^+({\rm or~}\pi^+)$,  and  other specific $B_s^0$ decay backgrounds including $B_s^0\to J/\psi \eta'$, $\eta'\to \rho\gamma$, $B_s^0\to J/\psi \phi$, $\phi\to \pi^+\pi^-\pi^0$. The shape of the sum of the combinatorial and  $B^+\to J/\psi K^+(\pi^+)$ components is taken from the like-sign events. The shapes of the other components are taken from Monte Carlo simulation with their normalizations allowed to float.

 We perform a simultaneous unbinned likelihood fit to the $\pi^+\pi^-$ opposite-sign and sum of $\pi^+\pi^+$ and $\pi^-\pi^-$ like-sign event distributions.
The fit gives a $B_s^0$ mass of 5366.1$\pm$1.1 MeV in good agreement with the known mass of  5366.3$\pm$0.6  MeV, a  Gaussian width of 8.2$\pm$1.1 MeV, consistent with the expected mass resolution and 111$\pm$14 signal events
within $\pm$30 MeV of the $\Bs$ mass. The change in twice the natural logarithm of the fit likelihood when removing the $B_s^0$ signal component, shows that the signal has an equivalent of 12.8 standard deviations of significance. The like-sign di-pion yield correctly describes the shape and level of the background below the $\Bs$ signal peak, both in data and Monte Carlo simulations. There are also 23$\pm$9 $B^0\to J/\psi\pi^+\pi^-$ events.

Having established a clear signal, we perform certain checks to ascertain if the structure peaking near 980 MeV is a spin-0 object. Since the $\Bs$ is spinless, when it decays into a spin-1 $J/\psi$ and a spin-0 $f_0$, the decay angle of the $J/\psi$ should be distributed as 
 \mbox{$1-\cos^2\theta_{J/\psi}$}, where $\theta_{J/\psi}$ is the angle of the $\mu^+$ in the $J/\psi$ rest frame with respect to the $\Bs$ direction. The polarization angle, $\theta_{f_0}$, the angle of the $\pi^+$ in the $f_0$ rest frame with respect to the $\Bs$ direction,
should be uniformly distributed.  A simulation of the $J/\psi$ detection efficiency in these decays shows that it is approximately independent of $\cos\theta_{J/\psi}$. The acceptance for $f_0\to\pi^+\pi^-$ as a function of the $\pi^+$ decay angle shows an inefficiency of about 50\% at $\cos\theta_{f_0}=\pm1$ with respect to its value at $\cos\theta_{f_0}=0$. It is fit to a parabola  and the inefficiency corrected in what follows.

\begin{figure}[hbt]
\centering
\includegraphics[width=6.in]{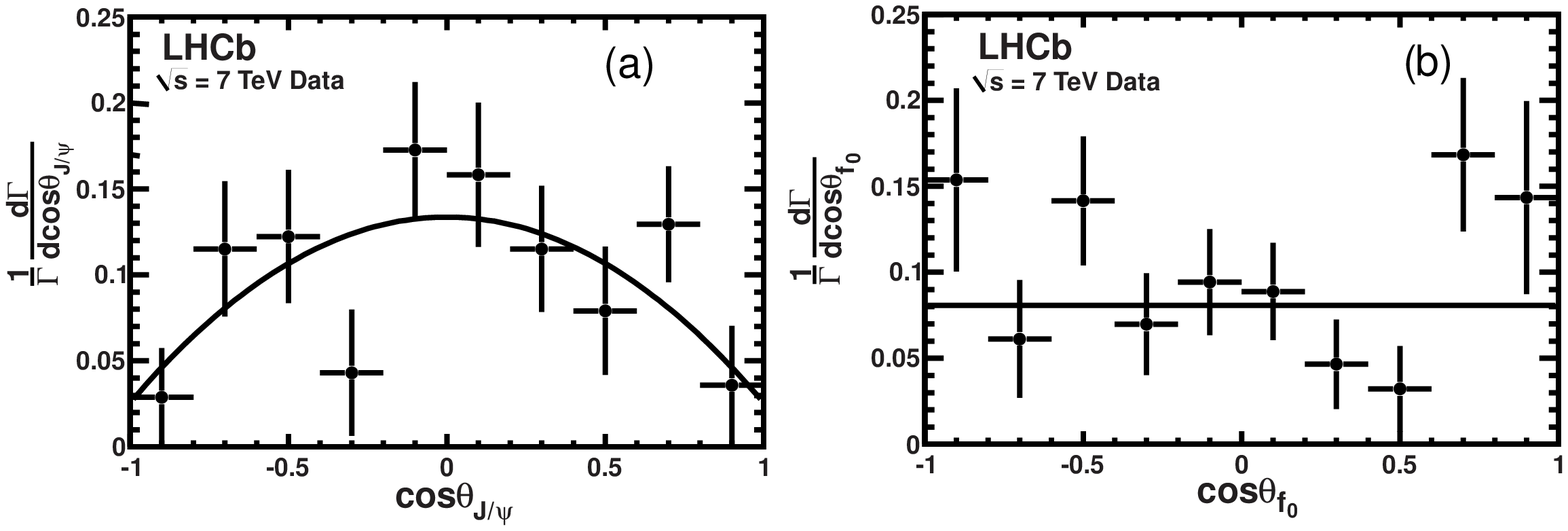}
\caption{Angular distributions of events within $\pm$30 MeV of the $\Bs$ mass and $\pm$90 MeV of the $f_0$ mass after like-sign background subtraction. (a) The cosine of the angle of the $\mu^+$ with respect to the $\Bs$ direction in the $J/\psi$ rest frame for $\Bs\to J/\psi \pi^+\pi^-$ decays. The data are fit with a function $f(\cos\theta_{J/\psi})=1-\alpha\cos^2\theta_{J/\psi}$. (b) The cosine of the angle of the $\pi^+$ with respect to the $\Bs$ direction in the di-pion rest frame for $\Bs\to J/\psi\pi^+\pi^-$ decays. The data are fit with a flat line.
}
\label{hel_sub}
\end{figure}

The like-sign background subtracted
$J/\psi$ helicity distribution is fit to a $1-\alpha\cos^2\theta_{J/\psi}$
function as shown in Fig.~\ref{hel_sub}(a). The fit gives $\alpha= 0.81\pm0.21$
consistent with a longitudinally polarized $J/\psi$ (spin perpendicular to its momentum) and a spin-0 $f_0$ meson. The $\chi^2$ of the fit is 10.3 for 8 degrees of freedom.
Similarly, we subtract the like-sign background and fit the efficiency corrected $\pi^+\pi^-$ helicity distribution to a constant function as shown in Fig.~\ref{hel_sub}(b). The fit has a $\chi^2$/ndof equal to 15.9/9, still consistent with a uniform distribution as expected for a spinless particle.

\begin{figure}[!hbt]
\centering
\includegraphics[width=5.in]{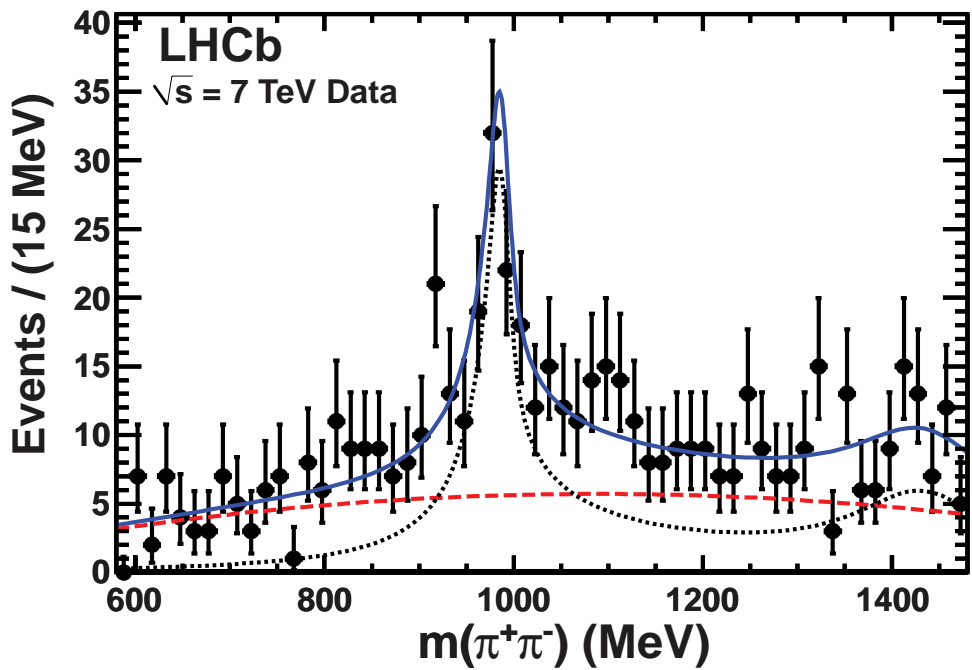}
\caption{The invariant mass of $\pi^+\pi^-$ combinations when the $J/\psi\pi^+\pi^-$  is required to be within $\pm$30 MeV of the $\Bs$ mass. The dashed curve is the like-sign background that is taken from the data both in shape and absolute normalization. The dotted curve is the result of the fit using Eq.~\ref{eq:fit} and the solid curve the total.
}
\label{fitnew_res}
\end{figure}

To view the spectrum of $\pi^+\pi^-$ masses, between 580 and 1480 MeV, in the $J/\psi\pi^+\pi^-$ final state we select events within $\pm$30 MeV of the $B_s^0$ and plot the invariant mass spectrum in Fig.~\ref{fitnew_res}.
The data show a strong peak near 980 MeV and an excess of events above the like-sign background extending up to 1500 MeV. Our mass spectrum is similar in shape to those seen previously in studies of the S-wave $\pi^+\pi^-$ system with $s\overline{s}$ quarks in the initial state \cite{BaBar-Swave, BES, E791}.
To establish a value for $R_{f_0/\phi}$ requires fitting the shape of the $f_0$ resonance.  Simulation shows that our acceptance is independent of the $\pi^+\pi^-$ mass, and we choose an interval between 580 and 1480 MeV.  Guidance is given by the BES collaboration who fit the spectrum in $J/\psi\to\phi\pi^+\pi^-$ decays \cite{BES}. We include here the $f_0(980)$ and $f_0(1370)$ resonances, though other final states may be present, for example the $f_2(1270)$ a $2^{++}$ state \cite{BaBar-Swave, BES}; it will take much larger statistics to sort out the higher mass states.
We use a coupled-channel Breit-Wigner amplitude (Flatt\'e) for the $f_0(980)$ resonance \cite{Flatte} and a
Breit-Wigner shape (BW) for the higher mass $f_0(1370)$. Defining $m$ as the $\pi^+\pi^-$ invariant mass,
the mass distribution is fit with a function involving the square of the interfering amplitudes
\begin{equation}
\label{eq:fit}
\left|A(m)\right|^2 =N_0mp(m)q(m)\left |{\rm Flatt\mbox{\'e}}[f_0(980)] + A_1\exp^{(i\delta)}{\rm BW}[f_0(1370)]\right|^2,
\end{equation}
where $N_0$ is a normalization constant, $p(m)$ is the momentum of the $\pi^+$, $q(m)$ the momentum of the    $J/\psi$ in the $\pi^+\pi^-$ rest-frame, and $\delta$ is the relative phase between the two components. The Flatt\'e amplitude is defined as
\begin{equation}
{\rm Flatt\mbox{\'e}}(m)=\frac{1}{m_0^2-m^2-im_0(g_1\rho_{\pi\pi}+g_2\rho_{KK})}~,
\end{equation}
where $m_0$ refers to the mass of the $f_0(980)$ and $\rho_{\pi\pi}$ and $\rho_{KK}$  are Lorentz invariant phase space factors equal to $2p(m)/m$ for $\rho_{\pi\pi}$. The $g_2\rho_{KK}$ term accounts for the opening of the kaon threshold. Here $\rho_{KK}=2p_K(m)/m$ where $p_K(m)$ is the momentum a kaon would have in the $\pi^+\pi^-$ rest-frame. It is taken as an imaginary number when $m$ is less than twice the kaon mass. We use  $m_0g_1=0.165\pm 0.018$ GeV$^2$, and $g_2/g_1 = 4.21\pm 0.33$ as determined by BES \cite{BES}.

The $f_0(1370)$ mass and width values used here are 1434$\pm$20 MeV, and 172$\pm$33 MeV from an analysis by E791 \cite{E791}. We fix the central values of these masses and widths  in the fit, as well as $m_0g_1$ and the $g_2/g_1$ ratio for the $f_0(980)$ amplitude. The mass resolution is incorporated as a Gaussian convolution in the fit as a function of $\pi^+\pi^-$ mass. It has an r.m.s. of 5.4 MeV at 980 MeV.
We fit both the opposite-sign and like-sign distributions simultaneously.
The results of the fit are shown in Fig.~\ref{fitnew_res}.  The $\chi^2$/ndof is 44/56. We find an $f_0(980)$ mass value of 972$\pm$25 MeV. There are 265$\pm$26 events above background in the extended mass region, of which $(64^{+10}_{-~6})$\% are associated with the $f_0(980)$, $(12\pm 4)$\% are ascribed to the $f_0(1370)$ and $(24^{+2}_{-6})$\% are from interference. The fit determines $\delta=61\pm36^{\circ}$. The fit fraction is defined as the integral of a single component divided by the coherent sum of all components. The $f_0(980)$ yield is 169$^{+31}_{-21}$ events. The lower mass cutoff of the fit region loses 1\% of the $f_0(980)$ events. The change in twice the log likelihood of the fit when removing the $f_0(980)$ component shows that it has an equivalent of 12.5 standard deviations of significance. 

Using the 169 $f_0$  events from $J/\psi\pi^+\pi^-$,  and the  635 $\phi$ events from $J/\psi K^+K^-$, correcting by the relative efficiency, and ignoring a possible small S-wave contribution under the $\phi$ peak \cite{CDF-upper},  yields
\begin{equation}
R_{f_0/\phi}\equiv\frac{\Gamma(B_s^0\to J/\psi f_0,~f_0\to \pi^+\pi^-)}{\Gamma(B_s^0\to J/\psi \phi,~\phi\to K^+K^-)}=0.252^{+0.046+0.027}_{-0.032-0.033}~.
\end{equation}
Here and throughout this Letter whenever two uncertainties are quoted the first is statistical and the second is systematic. 
This value of $R_{f/\phi}$
depends on the decay amplitudes used to fit the $\pi^+\pi^-$ mass distribution and could change
with different assumptions. To check the robustness of this result,  an incoherent phase space background is added to the above fit function. The number of signal $f_0(980)$ events is decreased by 7.3\%. If we leave the $f_0(1370)$ out of this fit, the original $f_0(980)$ yield is decreased by 6.5\%. The larger number of these two numbers is included in the systematic uncertainty. The BES collaboration also included a $\sigma$ resonance in their fit to the $\pi^+\pi^-$ mass spectrum in $J/\psi\to \phi \pi^+\pi^-$ decays \cite{BES}. We do not find it necessary to add this component to the fit.

The systematic uncertainty has several contributions listed in Table~\ref{tab:syserr}. There is an uncertainty due to our kaon and pion identification. The identification efficiency is measured with respect to the Monte Carlo simulation using samples of $D^{*+}\to \pi^+ D^0$, $D^0\to K^-\pi^+$ events for kaons, and samples of $K_S^0\to\pi^+\pi^-$ decays for pions. The correction to $R_{f_0/\phi}$ is 0.947$\pm$0.009.  This correction is already included in the efficiencies quoted above, and the 1\% systematic uncertainty is assigned for the relative particle identification efficiencies.

The efficiency for detecting $\phi\to K^+K^-$ versus a $\pi^+\pi^-$ pair is measured using $D^+$ meson decays into $\phi\pi^+$ and $K^-\pi^+\pi^+$ in a sample of semileptonic $B$ decays where $B\to D^+ X \mu^-\overline{\nu}$ \cite{CLEO-c-KKpi}. The simulation underestimates the $\phi$ efficiency relative to the $\pi^+\pi^-$ efficiency by (6$\pm$9)\%, so we take 9\% as the systematic error.

Besides the sources of uncertainty discussed above, there is a variation due to varying the parameters of the two resonant contributions.  We also include an uncertainty for a mass dependent efficiency as a function of $\pi^+\pi^-$ mass by changing the acceptance function from flat to linear and found that the $f_0$ yield changed by 2.3\%. The difference $\Delta\Gamma/\Gamma$ between $CP$ even and $CP$ odd $B_s$ eigenstates is taken as 0.088. Ignoring this difference results in less than a 1\% change in the relative efficiency. 
\begin{table}[!htb]
\centering
\caption{Relative systematic uncertainties on $R_{f_0/\phi}$ (\%). Both negative and positive changes resulting from the parameter variations are indicated in separate columns.}
\label{tab:syserr}
\begin{tabular}{lcc}
\hline\hline
Parameter & Negative change & Positive change\\\hline
$f_0(1370)$ mass & $0.3$ & 1.9 \\
$f_0(1370)$ width & $2.3$ &  2.6 \\
$\pi^+\pi^-$ mass dependent efficiency& 2.3 & 2.3\\
$m_0g_1$ & $4.2$ & 3.6\\
$g_2/g_1$ & $0.7$ & 0.7\\
Addition of non-resonant $\pi^+\pi^-$ & $7.3$ & 0\\
MC statistics (efficiency ratio) & $2.3$ & 2.3 \\
$\Bs$ $p_T$ distribution & 0.5&0.5\\
$\Bs$ mass resolution & 0.5 & 0.5 \\
PID efficiency & $1.0$& 1.0\\
$\phi$ detection &$9.0$ & 9.0\\\hline
Total & 13.1 & 10.8\\
\hline
\end{tabular}
\end{table}

In order to give a model independent result we also quote the fraction, $R'$, in the interval
$\pm$90 MeV around 980 MeV, corresponding to approximately two full-widths, where there are 111$\pm$14 events. Then
\begin{equation}
R'\equiv\frac{\Gamma \left(B_s^0\to J/\psi\pi^+\pi^-,~\left| m(\pi^+\pi^-)-980~{\rm MeV}\right|<90~{\rm MeV}\right)}{\Gamma(B_s^0\to J/\psi \phi,~\phi\to K^+K^-)}=0.162\pm 0.022\pm 0.016~.
\end{equation}
This ratio is based on the fit to the $B_s^0$ mass distribution and does not have any uncertainties related to the fit of the $\pi^+\pi^-$ mass distribution. Based on our fits to the $\pi^+\pi^-$ mass distribution, there are negligible contributions from any other signal components than the $f_0(980)$ in this interval.

The original estimate from Stone and Zhang was $R_{f_0/\phi}$ = 0.20 \cite{Stone-Zhang}. More recent predictions have been summarized by Stone \cite{S-waves} and have a rather wide range from 0.07 to 0.50.


\section{Conclusions}
Based on the polarization and rate estimates described above,  the first observation of 
a new CP-odd eigenstate decay mode of the $B_s^0$ meson into $J/\psi f_0(980)$ has been made.
Using a fit including two interfering resonances, the $f_0(980)$ and $f_0(1370)$, the ratio to $J/\psi \phi$ production is measured as
\begin{equation}
R_{f_0/\phi}\equiv\frac{\Gamma(B_s^0\to J/\psi f_0,~f_0\to \pi^+\pi^-)}{\Gamma(B_s^0\to J/\psi \phi,~\phi\to K^+K^-)}=0.252^{+0.046+0.027}_{-0.032-0.033}~.
\end{equation}
By selecting events within $\pm$90 MeV of the $f_0(980)$ mass the ratio becomes
$R'=0.162\pm 0.022\pm 0.016~.$

The events around the $f_0(980)$ mass are large enough in rate and have small enough backgrounds that they could be used to measure $\beta_s$ without angular analysis. It may also be possible to use other data in the $\pi^+\pi^-$ mass region above the $f_0(980)$ for this purpose if they turn out to be dominated by S-wave.

\section{Acknowledgments}
We express our gratitude to our colleagues in the CERN accelerator departments for the excellent performance of the LHC.
We thank the technical and administrative staff at CERN and at the LHCb institutes, and acknowledge support from the National Agencies: CAPES, CNPq, FAPERJ and FINEP (Brazil); CERN; NSFC (China); CNRS/IN2P3 (France); BMBF, DFG, HGF and MPG (Germany); SFI (Ireland); INFN (Italy); FOM and NWO (Netherlands); SCSR (Poland); ANCS (Romania); MinES of Russia and Rosatom (Russia); MICINN, XUNGAL and GENCAT (Spain); SNSF and SER (Switzerland); NAS Ukraine (Ukraine); STFC (United Kingdom); NSF (USA). We also acknowledge the support received from the ERC under FP7 and the R\'egion Auvergne.


\begin{thebibliography}{99}
\bibitem{CDF-upper}
The CDF collaboration limits the S-wave contribution within $\pm$10 MeV of the $\phi$ mass to 6.7\% at 95\%
confidence level.  Their best fit value is 2\% \cite{TevatronCDF}.
\end{thebibliography}
\end{document}